\documentclass{aa}

\usepackage{graphics}

\def\kms{\hbox{\,km\,s$^{-1}$}}

\begin{document}

\thesaurus{06(08.05.3; 08.09.2 $\eta$ Car; 08.13.2; 09.02.1;
09.10.1)}

\title{High velocity structures in, and the X-ray emission
from\\
the LBV nebula around $\eta$\,Carinae}

\author {
K.\ Weis \inst{1,2,3}\thanks{Visiting Astronomer, Cerro Tololo
Inter-American Observatory, National Optical Astronomy
Observatories, operated by the Association of Universities for
Research in Astronomy, Inc., under contract with the National
Science Foundation.} \and W.J.\ Duschl \inst{1,2} \and D.J.\
Bomans \inst{4} }

\offprints{K.\ Weis, email: kweis@ita.uni-heidelberg.de}

\mail{K.\ Weis, Institut f\"ur Theoretische Astrophysik,
Tiergartenstr. 15, 69121 Heidelberg, Germany}

\institute{ Institut f\"ur Theoretische Astrophysik,
Tiergartenstr. 15, 69121 Heidelberg, Germany \and
Max-Planck-Institut f\"ur Radioastronomie, Auf dem H\"ugel 69,
53121 Bonn, Germany \and University of Illinois, Department of
Astronomy, 1002 W. Green Street, Urbana, IL 61801, USA \and
Astronomisches Institut, Ruhr-Universit\"at Bochum,
Universit\"atsstr. 150, 44780 Bochum, Germany }

\date{received; accepted}

\maketitle

\begin{abstract}

The Luminous Blue Variable star $\eta$ Carinae is one of the most
massive stars known. It underwent a giant eruption in 1843 in
which the Homunculus nebula was created. ROSAT and ASCA data
indicate the existence of a hard and a soft X-ray component which
appear to be spatially distinct: a softer diffuse shell of the
nebula around $\eta$ Carinae and a harder point-like source
centered on the star $\eta$ Car. Astonishingly the morphology of
the X-ray emission is very different from the optical appearance
of the nebula. We present a comparative analysis of optical
morphology, the kinematics, and the diffuse soft X-ray structure
of the nebula around $\eta$ Carinae. Our kinematic analysis of the
nebula shows extremely high expansion velocities. We find a strong
correlation between the X-ray emission and the knots in the nebula
and the largest velocities, i.e. the X-ray morphology of the
nebula around $\eta$ Carinae is determined by the interaction
between material streaming away from $\eta$ Car and the ambient
medium.


\end{abstract}

\section{Introduction}

\subsection{$\eta$ Carinae as a Luminous Blue Variable Star}

The star $\eta$ Carinae, embedded in the large Carina H\,{\sc ii}
complex (the Carina nebula) is still one of
the most unique stellar objects in our
Galaxy. Known as a variable star for centuries, it brightened up
to $-$1$^{\rm m}$ around 1843 (Herschel 1847, Innes 1903, van
Genderen \& Th\'e 1984, Viotti 1995, Humphreys et al.\ 1999) and
drastically decreased its brightness by more than  7$^{\rm m}$
within just 20 years. While the brightness declined, discussions
on the nature of this outburst started. A new hint on the origin
occurred a century after the burst when nearly simultaneously
Gaviola (1946, 1950) and Thackeray (1949, 1950) discovered a
nebula around $\eta$ Car. Since the appearance of the nebula on
the first image was of man-like shape, Gaviola named it the {\it
Homunculus}. Today $\eta$ Car is classified as a {\it Luminous
Blue Variable\/} (LBV).

The most massive stars (with zero-age main-sequence masses $\ge
50$\,M$_{\sun}$) start as main-sequence O stars and evolve
quickly into supergiants. While they cross the
Hertzsprung-Russell diagram (HRD) towards the red they seem to
enter an unstable phase at an age of roughly 3 10$^6$ years
(Langer et al.\ 1994). This phase, the LBV phase, goes along with
a very high mass loss (rates up to
several\,$10^{-4}$\,M$_{\sun}\,{\rm yr}^{-1}$). The strong
stellar wind and possible giant eruptions where parts of the star's
envelope are peeled off, often form circumstellar nebulae, the so
called {\it LBV nebulae\/} (e.g., Nota et al.\ 1995). The LBV phase
starts when the stars reach the {\it Humphreys-Davidson limit\/}
(Humphreys \& Davidson 1979, 1994) in the HRD. This empirical
limit marks the red end of the distribution of very luminous
supergiants (Humphreys 1978, 1979 and Humphreys \& Davidson
1979). The most luminous, massive stars do not evolve into red
supergiants but instead reverse their evolution towards the blue
supergiant part in the HRD when they reach the Humphreys-Davidson
limit, they turn into LBVs.

With a mass of $M \sim 120$\,M$_{\sun}$ and a luminosity of
$L\sim 10^{6.7}$\,L$_{\sun}$ (Humphreys \& Davidson 1994,
Davidson \& Humphreys 1997) $\eta$ Car is the most massive member
of the LBV class known. In the light of recent discussions about
$\eta$ Car being a binary (Damineli 1996, Damineli et al.\ 1997,
Stahl \& Damineli 1998), the masses of the two components  would
be between 65 and 70\,M$_{\sun}$ each. Both components would
therefore still be among the most massive  stars. The Homunculus
nebula is believed to have formed during a giant eruption---the
brightness increase of 1843. Images taken with the {\it Hubble
Space Telescope\/} (HST) revealed the bipolar structure of the
Homunculus, consisting of two lobes each about $8-9\arcsec$ in
diameter (e.g., Morse et al.\ 1998). The original man-like shape
seen around 1950 represents only the brightest emission of the
bipolar lobes in the central region. Beside the two lobe structure
an equatorial disk was found already through ground-based
observations (Duschl et al.\ 1995). The deepest HST pictures (200s
in the F656N-filter) show a larger amount of very complex
filamentary structures like knots, arcs and strings (Weis et al.\
1999) of which the sizes vary between fractions of arcseconds and
several arcseconds. These structures extend much further out than
the bipolar Homunculus, up to a distance of at least 30\arcsec\
and form the outer nebula, the so-called {\it outer ejecta\/}
(Weis 2000, Weis \& Duschl, in prep.). The LBV nebula around
$\eta$ Carinae therefore consists of the inner bipolar Homunculus
($\sim$ 17\arcsec\ in diameters) and the filemantary outer ejecta
(up to 60\arcsec\ across).

Kinematic analysis of the nebula around $\eta$ Carinae detect astonishingly
high expansion velocities, especially in the outer ejecta. In several
publications radial
velocities of the bipolar central and the outer nebula as high as
1000\kms\ were reported (Meaburn et al.\ 1987, 1993, 1996, Hillier
\& Allen 1992, Weis et al.\ 1999, Weis \& Duschl, in prep.).
Similar velocities were derived from proper motion measurements
(Walborn 1976, Walborn et al.\ 1978, Walborn \& Blanco 1988, Currie
et al.\ 1996).

$\eta$ Car was also found to emit in the X-ray regime, with the
star being an individual point source and the dominant contributor
of hard X-rays while an extended  soft X-ray emission was observed
from the nebula around $\eta$ Carinae.

\subsection{Chronology of the X-ray observations}

X-ray emission of the $\eta$ Car complex was detected by Hill
(1972), observing with a proportional counter system aboard a {\it
Terrier-Sandhawk} rocket. Observations with {\it Ariel 5} (Seward
et al.\ 1976), {\it OSO 8} (Becker et al.\ 1976, Bunner 1978) and
{\it Uhuru} (Forman et al.\ 1978) followed. With the launch of the
{\it Einstein Observatory} the resolution of X-ray images was
pushed down to a few arcseconds with the high resolution imager
(HRI). Therefore for the first time the X-ray sources in the
Carina complex could be resolved into several stars and a soft
X-ray emitting diffuse component of the larger Carina H\,{\sc ii}
complex (Seward et al.\ 1979, Seward \& Chlebowski 1982). $\eta$
Car itself was just one extended source. The extended soft X-rays
compared to the optical data, showed their origin in the nebula
around $\eta$ Carinae (Seward et al.\ 1979). The authors excluded
a supernova as formation mechanism for the X-radiation and
proposed that they formed through a blast wave from the $\eta$ Car
outburst. The Einstein observations also showed that there was a
hard X-ray source located somewhere in the Homunculus.

The  first detailed analysis of the X-rays from $\eta$ Car and its
nebula, with a longer Einstein observation, as well as a
comparison of the X-ray with the optical image was done in 1984
(Chlebowski et al.). This first overlay already identified the
most intense X-ray emission with structures in the outer ejecta,
namely the {\it S condensation\/}, the {\it W arc\/} and the {\it
E condensation\/} (for the notation of the knots see Walborn
1976). The observed X-rays result mainly from the outer shell of
$\eta$ Car and are soft in nature. In addition the harder central
source (not resolved with the Einstein {\it Imaging Propotional
Counter\/}, IPC) was tentatively identified with the star $\eta$
Car itself. The measurements of the {\it Ginga} satellite
sensitive to hard X-rays ($2-37$\,keV) prove the existence of a
harder source with a temperature of 4.1 keV in the Carina nebula
(Koyama et al.\ 1990).

ROSAT improved the resolution and sensitivity in the soft X-ray
band even further. $\eta$ Car was observed with ROSAT in both the
{\it Position Sensitive Proportional Counter\/} (PSPC; spatial
resolution $\sim 25\arcsec$)  and in the {\it High Resolution
Imager\/} (HRI; spatial resolution $\sim 5\arcsec$). Several
discussions on the ROSAT data prove now the existence of a harder
and softer X-ray component of $\eta$ Car and its nebula (c.f.,
Corcoran et al.\ 1994, 1995, 1996, 1997), which are spatially
distinct: a softer diffuse shell of the nebula and a harder
point-like source centered on $\eta$ Car. In addition it was found
that the hard  X-ray source shows variability and a pointlike
character (Corcoran et al.\ 1995).

Analogous results were achieved with the {\it Advanced Satellite
for Cosmology and Astrophysics} (ASCA; Tsuboi et al.\ 1997,
Corcoran et al.\ 1998). Due to the good spectral resolution of
ASCA it was also possible to conclude that the `X-ray variability
does not involve measureable changes to the spectral shape of the
emission' (Corcoran et al.\ 1998).

\subsection{Structure of the paper}

In this paper, we present the results of a comparative study of
images from the HST, high-resolution long-slit echelle spectra
and X-ray data. We will give an interpretation of the origin of
the X-rays and explain the different morphological appearance of
the optical and the X-ray emission of the nebula around $\eta$ Car.

\section{Observation and data reduction}

\begin{figure}

{\resizebox{\hsize}{!}{\includegraphics{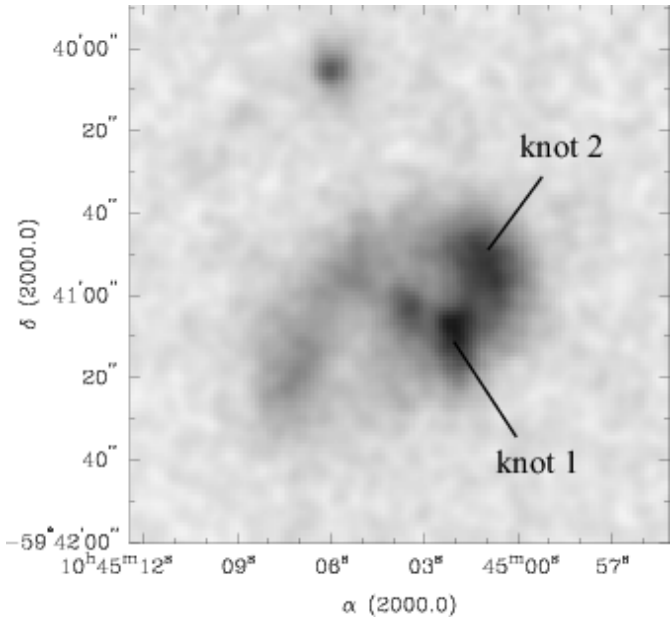}}}

\caption{ROSAT HRI image of the full usable integration time of
$\eta$ Car and its nebula. The field of view is about 2\arcmin
$\times$2\arcmin. The image was smoothed with a Gaussian filter to
an effective resolution of 5\farcs5. The marked knots 1 and 2 are
defined in the text.}

\label{fig:hrixray}

\end{figure}

\begin{figure}

\centerline{\resizebox{0.9\hsize}{!}{\includegraphics{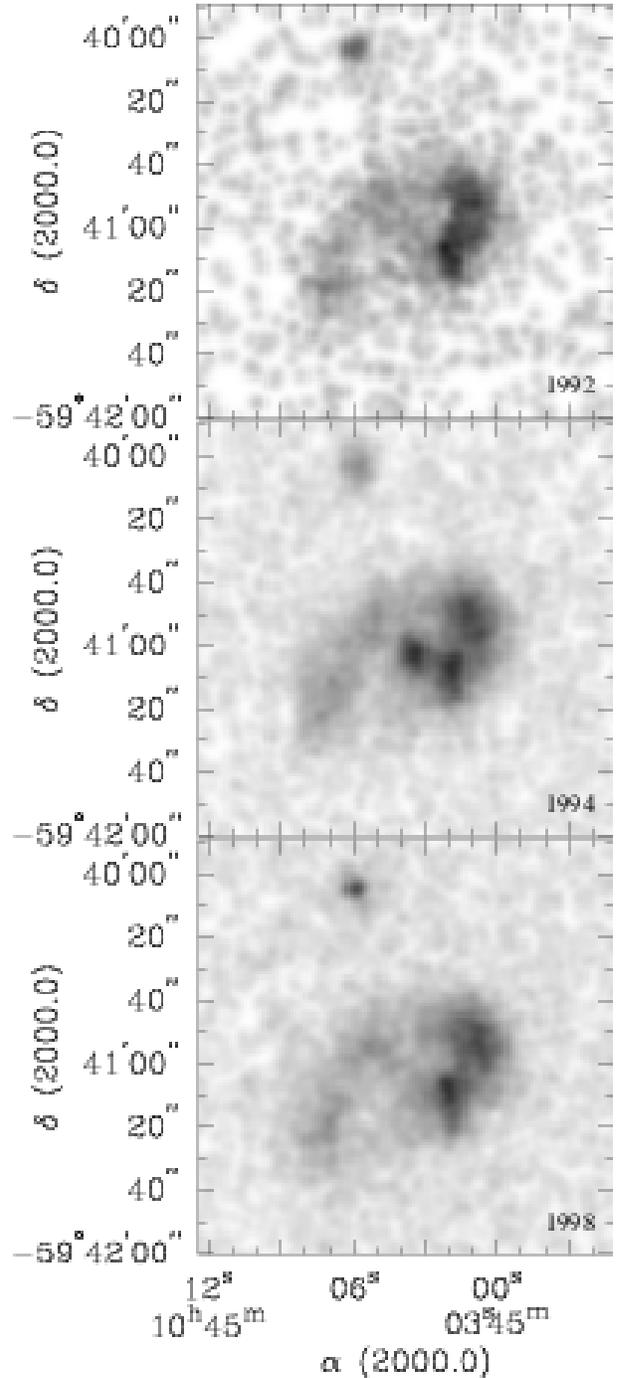}}}

\caption{HRI images of the $\eta$ Carinae nebula during the three
observing intervalls which are long enough to show the X-ray
nebula (see table \ref{tab:hridata}). The three panels show the
images generated from the 1992 (upper panel), 1994 (middle panel)
1998 (lower panel) data sets.} \label{fig:hrioo}

\end{figure}

\subsection{X-ray data}

For the analysis of the X-ray emission from the LBV nebula around
$\eta$ Car we made use of archived data from the {\it
R\"ontgensatellit (ROSAT)}. The ROSAT satellite was sensitive to
X-ray emission between 0.1 and 2.4 keV. Images of the $\eta$ Car
region were taken with the PSPC as well as the HRI. We used
mainly the high spatial resolution data taken with the HRI (first
HRI images published by Corcoran et al.\ 1996) since the small
X-ray nebula around $\eta$ Car is barely resolved in the PSPC data
(Corcoran et al.\ 1994). All exposures available for the Carina
region (see Tab.\ \ref{tab:hridata}) were retrieved from the {\it
Max-Planck-Institut f\"ur Extraterrestrische Physik\/} (MPE) ROSAT
data center.

\begin{table*}
\caption[]{ROSAT HRI Observations of $\eta$ Car; the column offset gives
the offset of the image center from $\eta$ Carinae.}
\begin{flushleft}
\begin{tabular}{cclrll}
\hline
pointing & Obs. interval & P.I. & exposure time [s] & offset [\arcmin] &
comment\\
\hline
rh150037n00 & 900727-900729 & Puls        &  3351 & 7.135 & \\
rh900385n00 & 920731-920802 & Schmitt     & 11527 & 0.31   & source off\\
rh900385a01 & 940106-940106 & Schmitt     &   522 & 0.31  & \\
rh900385a02 & 940721-940729 & Schmitt     & 40555 & 0.31  & source on\\
rh900644n00 & 960813-960813 & Corcoran    &  1720 & 0.31  & \\
rh202331n00 & 971223-980210 & Corcoran    & 47095 & 0.31  & source off\\
\hline
\end {tabular}
\end{flushleft}
\label{tab:hridata}
\end{table*}

Data reduction was performed with IRAF\footnote{IRAF is
distributed by the National Optical Astronomy Observatories which
is operated by AURA, Inc. under cooperative agreement with the
NSF.}/PROS\footnote{PROS is developed, distributed, and
maintained by the Smithonian Astrophysical Observatory, under
partial support from NASA contracts NAS5-30934 and NAS8-30751.}.
Three individual HRI pointings were long enough ($> 5$\,ksec) to
contain significant information about the diffuse emission around
$\eta$ Car.  First we screened the data for periods with
excessively high background. Since highest possible spatial
resolution was essential for our work, as next step we checked for
errors in the aspect solution, using methods similar to the
methods described in Harris et al.\ (1998). To ensure a stable
pointing, we excluded observing time with less than three guide
stars present. The total usable integration time adds up to
95\,ksec. Next we produced images of the individual observational
intervals (OBIs) and re-centered these individual images. Shifts
were generally small but not negligible, between 1\arcsec\ and
4\arcsec. The count rates in the brightest X-ray point source,
which could be used for positional shifts, was unfortunately too
low to split the OBIs even more into individual phase bins to
investigate residual effects of the spacecraft wobble. We chose a
blocking factor of 2, resulting in 1\arcsec\ per pixel and
smoothed the images slightly with a $\sigma=1.5$ pixel Gauss
filter. We updated the coordinates using the five X-ray point
sources coinciding with the stars HD 93205, HDE303308, CD$-59\degr
2635$, CD$-59\degr 2636$, and CD$-59\degr 2641$ in the Carina
nebula.  The coordinates of the stars were measured on a Digital
Sky Survey (DSS) image using the IRAF/GASP package. The resulting
positional accuracy of the X-ray images is of the order of
0\farcs5 relative to these stars. From the new photon file we
produced three images of the three observations in 1992, 1994 and
1998, according to the central source's high or low state. In
addition a total flux image (Fig.\ \ref{fig:hrixray}) was made,
combining all observations regardless of the state of the central
source. The images are shown in Fig.\ \ref{fig:hrioo}, in which
the upper panel gives the 1992 (low state), the middle one the
1994 (high state), and the lower one the 1998 (low state) data.

\subsection{Optical Imaging}

To compare the X-ray data with an optical image of the nebula
around $\eta$ Car, data were taken from the HST Archive at the
{\it Canadian Astronomy Data Centre\/} (CADC). Images of the {\it
Wide Field Planetary Camera\,2\/} (WFPC2) with the F656N (H$_{\alpha}$)
filter were retrieved\footnote{F656N -- program number: 5239;
P.I.: J.A. Westphal; dataset names: U2DH0101T \dots U2DH0106T}.
Their reduction, combination and cosmic-ray cleaning followed the
standard procedures recommended for WFPC2 data in IRAF. The
exposure times ranged from 0.11 to 200 seconds. The reduced and
mosaiced F656N image was used for an overlay with the total
integration time X-ray image (Fig. \ref{fig:hsthri}) and the
comparison between the X-ray and the optical emission. In
addition it was used to identify several knots in the outer
nebula, for which we took kinematic data.

\subsection{Long-slit echelle spectroscopy}

Kinematic information was obtained using long-slit echelle
spectroscopy. The data were taken by one of us (KW) with
the 4\,m Blanco telescope
at the {\it Cerro Tololo Inter-American Observatory\/} (CTIO) in
the long-slit mode. Wavelength selection was achieved by
inserting a post-slit H$_{\alpha}$ filter (6563/75\,\AA) and
replacing the cross-disperser by a flat mirror. We choose the 79\
l\,mm$^{-1}$ echelle grating, a slit-width of 250\,$\mu$m
($\widehat{=}\ 1\farcs 64$), which leads to an instrumental FWHM
at the H$_{\alpha}$ line of about 14\,km\,s$^{-1}$. The data were
recorded with the long focus red camera and the Tek2K4 CCD ($2048
\times 2048$). Here the pixel size was 0.08\,\AA\,pixel$^{-1}$
along the dispersion, and 0$\farcs$26\,pixel$^{-1}$ on the
spatial axis. Vignetting limited the slit length to
$\sim4^\prime$. During all observations the weather was not
photometric and the seeing was between $1\farcs5-2\arcsec$.
Thorium-Argon comparison lamp frames were taken for wavelength
calibration and geometric distortion correction.

\begin{figure}

{\resizebox{\hsize}{!}{\includegraphics{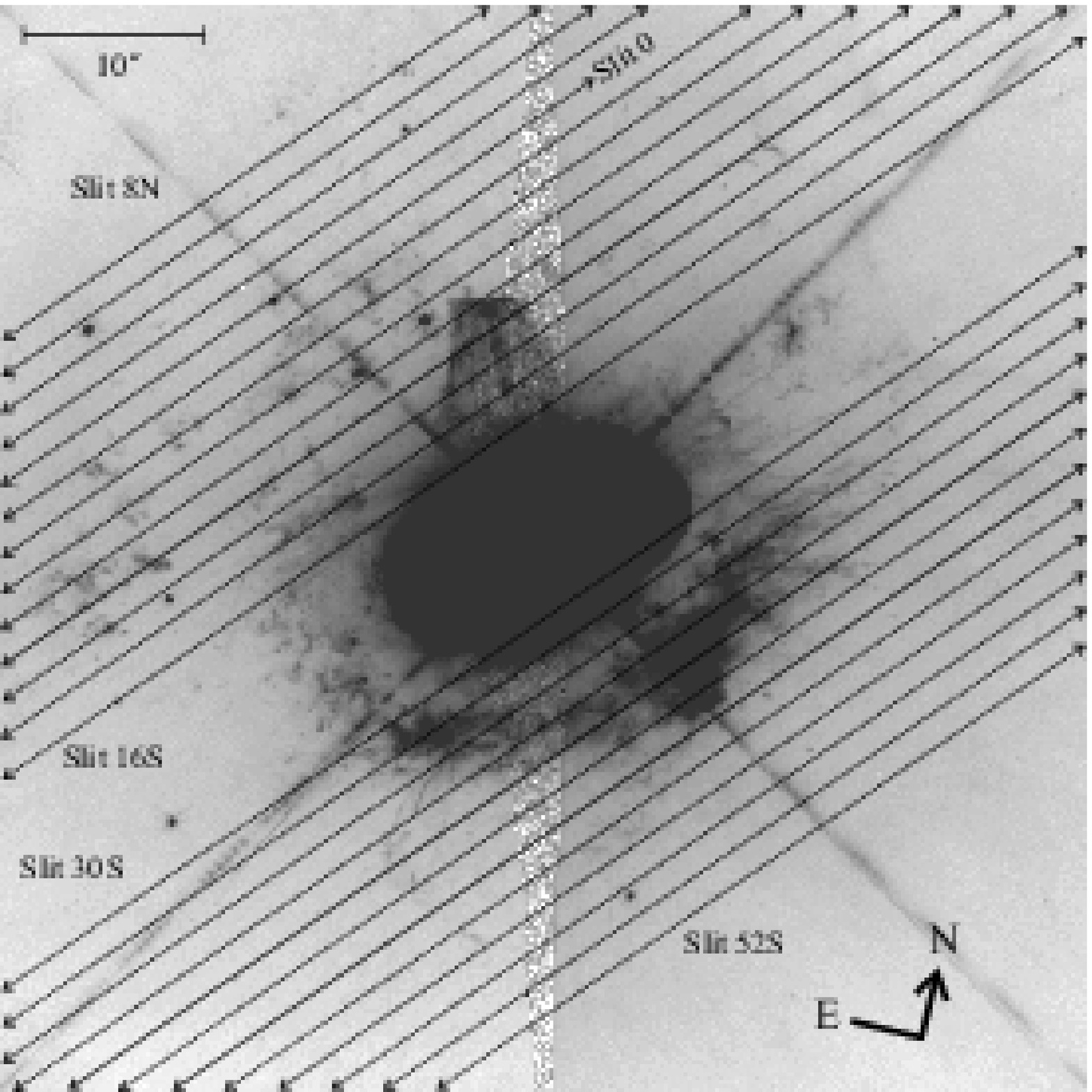}}}

\caption{This figure marks the position of the slits of the
echelle spectra with dashed and solid lines. The image shows a
60\arcsec $\times$ 60\arcsec\ section of the unrotated F656N HST
image from $\eta$ Carinae. A north-east vector indicates the
celestial orientation of the image. To avoid confusion, only some
necessary slit names are indicated. Slits drawn with solid rather
than dashed lines represent the positions of the slits for which
echellograms are given in Fig. \ref{fig:echellograms} }
\label{fig:slits}

\end{figure}

The whole dataset contains 31 slit positions, of which the
central 6 positions around $\eta$ Car could not be used, because
strong straylight from the dusty Homunculus and extended ghost
images did not allow us to extract reliable information from the
spectra. The individual observations were offset by 2\arcsec\
each from an offset star rather than $\eta$ Car itself, since the
position of $\eta$ Car is confused by strong emission of the
Homunculus. The orientation and position of the slits are
indicated in Fig. \ref{fig:slits}. With the mapping a field of
about 60\arcsec$\times$60\arcsec\ around $\eta$ Car was covered.
In Fig.\ \ref{fig:echellograms} we give three example spectra with
individual knots marked by numbers. The spectra cover a range of
80\,\AA\ centered on H$_\alpha$, thus they include also the
[N\,{\sc ii}] lines at 6548 and 6583\,\AA . The spatial extent is
1\farcm5. Usually the features are more prominent in the [N\,{\sc
ii}] lines than in H$_\alpha$ due to the CNO cycle processed
material in the nebula.

\begin{figure*}

{\resizebox{13.4cm}{!}{\includegraphics{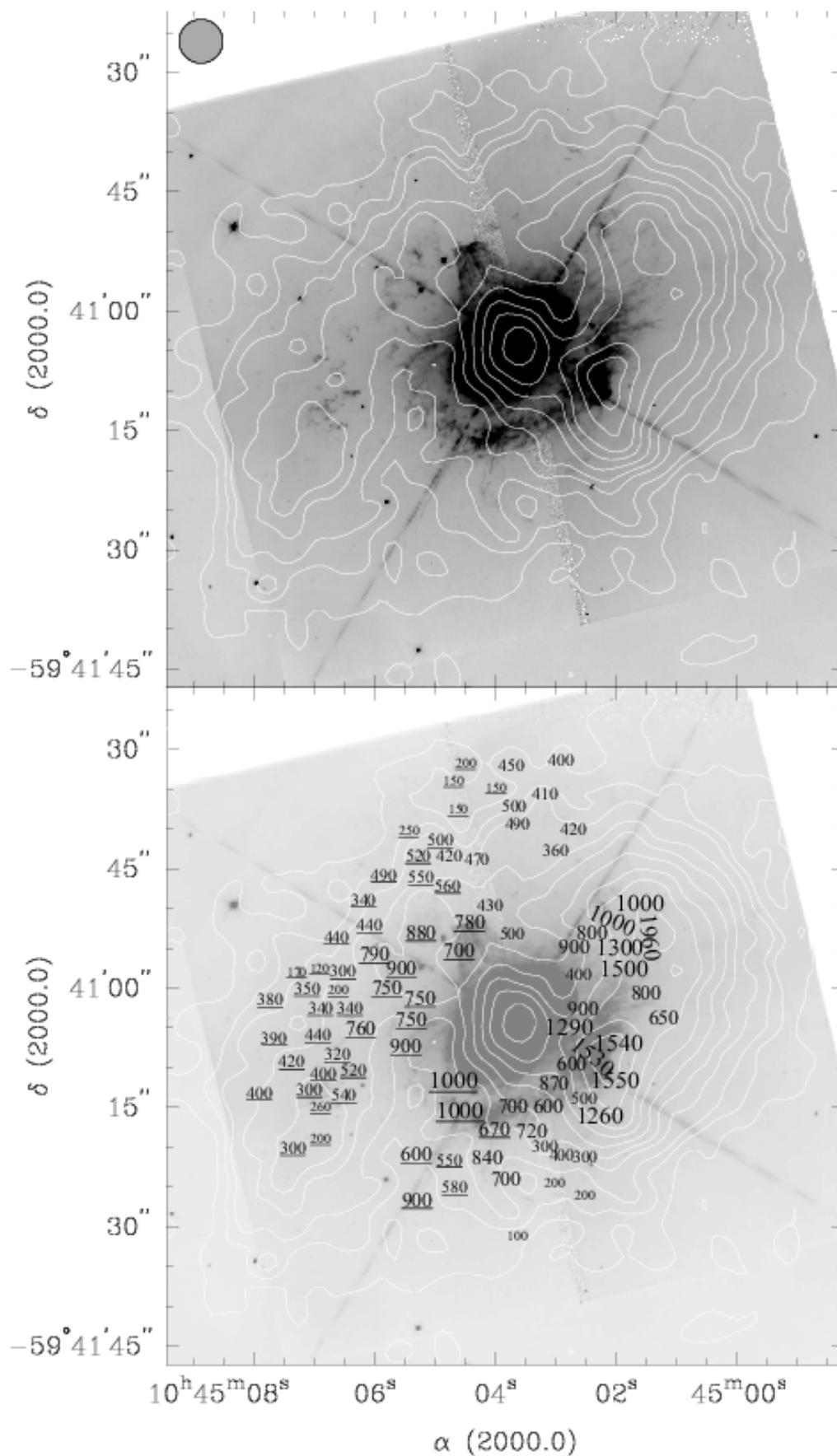}}}
\parbox[b]{41mm}{\caption{Overlay of the X-ray emission (contour lines) and
the HST image (grey scales): In the upper left corner of the top
panel the ROSAT HRI resolution (5\farcs5) is indicated. The lower
panel shows the same overlay (with decreased grey scale
intensity), with the measured radial velocities (in \kms) placed
at their respective positions. Underlined numbers indicate negative
velocities. The sizes of the characters increase with increasing
absolute velocities (see text). Due to their low surface
brightness, occasionally the clumps and knots for which
velocities are given, can be identified on the HST image but do
not show up clearly in this print.\label{fig:hsthri}}}

\end{figure*}

Figure \ref{fig:hsthri} shows the overlay of the X-ray emission
in contours onto the HST image, velocities (in \kms) represent
the kinematics in the lower panel. Clearly a lack of data can be
seen in the center, where the information of the 6 slit positions
is missing. A rotation of the slit placed the spectra to a
position angle of ${\rm PA} = 132\degr$, i.e. along the major
axis of the bipolar nebula. A more detailed description of the
observations and the spectra will be published separately (Weis \&
Duschl 2000, in preparation).

\section{X-rays from the $\eta$ Car region}\label{section:xray}

\subsection{The morphology of the $\eta$ Car X-ray nebula from ROSAT HRI data
\label{sect:xmorph}}

In this paper we concentrate on the X-ray emission from the
$\eta$ Car LBV nebula. The X-ray image with the highest spatial
resolution, the ROSAT HRI image, can be seen in Fig.
\ref{fig:hrixray}.

The X-ray nebula around $\eta$ Car extends much further out from
$\eta$ Car than the two lobes of the Homunculus (Fig.
\ref{fig:hsthri}), the {\it S Ridge\/} and the {\it N
Condensation\/} (for notations see Walborn 1976). Instead of the
bipolarity of the optical structure, the X-ray nebula consists of
a hook-shaped diffuse emission region, roughly encircling the
central point-like source. About one-third (position angles ${\rm
PA} \approx 210 - 330\degr$) of the loop is of high X-ray surface
brightness. At about ${\rm PA} \approx 330\degr$ the surface
brightness abruptly drops by a factor of three and stays at this
level for the range ${\rm PA} \approx 330 - 130\degr$; the
remainder is at an even weaker, barely detectable surface
brightness.

Several local maxima of the X-ray surface brightness are visible
in the X-ray nebula (Fig.\ \ref{fig:hrixray} and
\ref{fig:hsthri}). In the south-west the brightest maximum is
located ($\alpha \approx 10^{\rm h} 45^{\rm m} 2\fs0, \delta
\approx -59\degr 41\arcmin 08\arcsec$). In the following we will
refer to it as {\it knot 1\/}. It coincides roughly with the {\it S
condensation}. The X-ray maximum is approximately triangular, with a
corner pointing into the direction of the central source. A
fainter extension follows to the south. A second maximum (at
$\alpha \approx 10^{\rm h} 45^{\rm m} 1\fs4, \delta \approx
-59\degr 40\arcmin 53\arcsec$) peaks to the north-west of $\eta$ Car
({\it knot 2\/}), in a part of the nebula where no bright optical
counterpart but only low surface brightness clumps and knots can
be identified. Knot 2 in the X-ray image seems to consist of two
or three sub-entities.

\begin{figure*}

\resizebox{13.4cm}{!}{\includegraphics{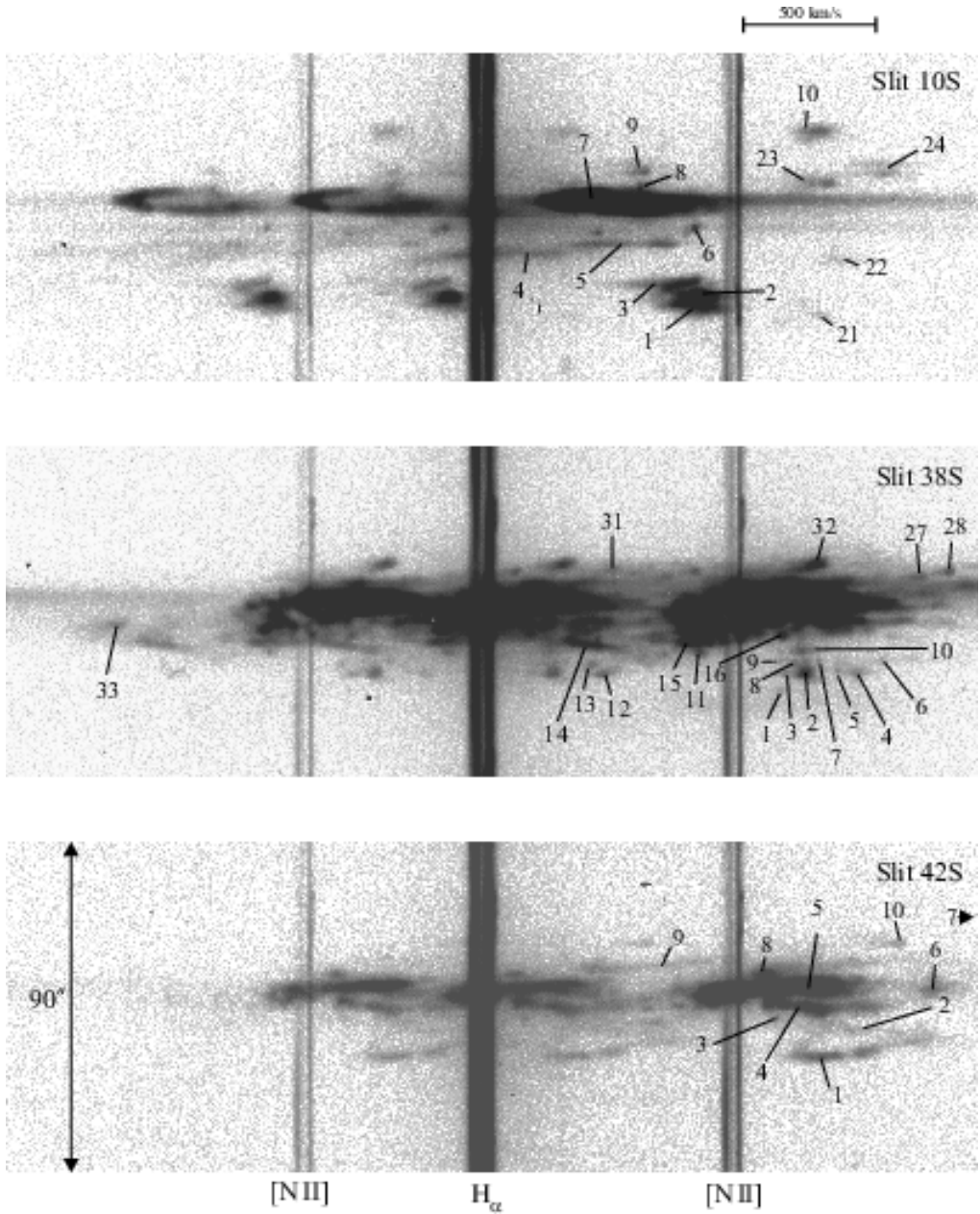}} \hfill
\parbox[b]{41mm}{
\caption{Three typical Echellograms of our mapping: They cover a
wavelength range of 80\,\AA\ each. Continuous split lines indicate
the expansion of the gas in the background nebula in the vicinity
of $\eta$ Car. Numerous knots and clumps are identified with a
broad range of velocities. Numbers mark individual
knots.\label{fig:echellograms}}}

\end{figure*}

Two local maxima in surface brightness are embedded in the low
surface brightness part of the X-ray nebula north and east of
$\eta$ Car (at $\alpha \approx 10^{\rm h} 45^{\rm m} 4\fs5;
\delta \approx -59\degr 41\arcmin 23\arcsec$ and at $\alpha
\approx 10^{\rm h} 45^{\rm m} 5\fs5, \delta \approx -59\degr
40\arcmin 58\arcsec$). The two maxima are less peaked and more
extended than the maxima in the high surface brightness part of
the $\eta$ Car X-ray nebula.

\begin{figure}

\resizebox{\hsize}{!}{\includegraphics{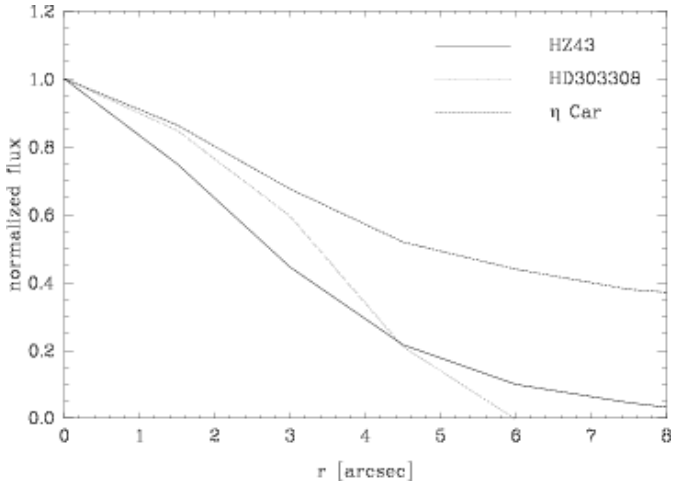}} \hfill

\caption{Radial X-ray brightness profiles of the central
source of $\eta$ Car and two comparison point sources, the most
nearby X-ray bright star in the  $\eta$ Car field (HD303308) and
the ROSAT standard point source HZ43. Note that the slope of the
$\eta$ Car profile is much shallower than the slope of the two
point sources. \label{fig:radial}}

\end{figure}

A closer look at the central point source indicates that it is
elongated (${\rm PA} \approx 30\degr$), while the other point
sources in the field are round. This excludes that the shape of
the central source is produced by residual uncertainties of OBI
centering or wobble. Deviation of the measured point spread
function of the HRI from the theoretical expected circular
symmetric shape occur (``meaty sources'') and are independent of
the wobble, but should be present in all sources with small
off-axis angles in a given exposure (Briel et al.\ 1994).
A radial plot of the central point source is shown in Fig.\
\ref{fig:radial} in comparison with the radial brightness
profiles of the nearest isolated star in the $\eta$ Car X-ray
image (HD303308), and the standard point source HZ43 (extracted
from calibration observations from the ROSAT archive).  While the
correction for the diffuse extended emission visible in the
radial profile of the central X-ray source of $\eta$ Car is a
matter of concern for this comparison, the slope of the fall-off
is shallower than the slope of the comparison point sources. The
elongation of the central source should therefore be intrinsic
and indicates a second extended emission component in addition to
$\eta$ Car itself. The orientation is consistent with it being
part of the equatorial disk, as defined by Duschl et al.\ (1995).
A first inspection of the $\eta$ Car data from the CHANDRA satellite
seem to support our observation, showing a point source and an
elongated halo, but these data definitely need a more careful
analysis.

When comparing the X-ray images produced from the three data sets
with sufficiently long exposure time (see Fig.\ \ref{fig:hrioo})
some differences can be discerned in the nebula itself between the
older low state image (1992 data set, Fig.\ \ref{fig:hrioo} upper
panel) and the high state image (1994 data set, Fig.\
\ref{fig:hrioo} middle panel), despite the lower signal-to-noise
ratio of the image at low-state. The shape of knot 1 appears
different, with the brightest spot being more distant from $\eta$
Car in the high state image which was taken at a later date. The
morphology of knot 2 appears different, too, but interpretation is
severely hampered by low count-rate in the low-state image. The
effective spatial resolution is the same in both images, the more
fuzzy looking appearance of HDE303308 (north of $\eta$ Car) is an
effect of the different brightness cuts of the images. The faint
nebula south-east of the central source seems to be connected with
diffuse emission visible at the position of the point source (at
low state).

\begin{figure}

{\resizebox{\hsize}{!}{\includegraphics{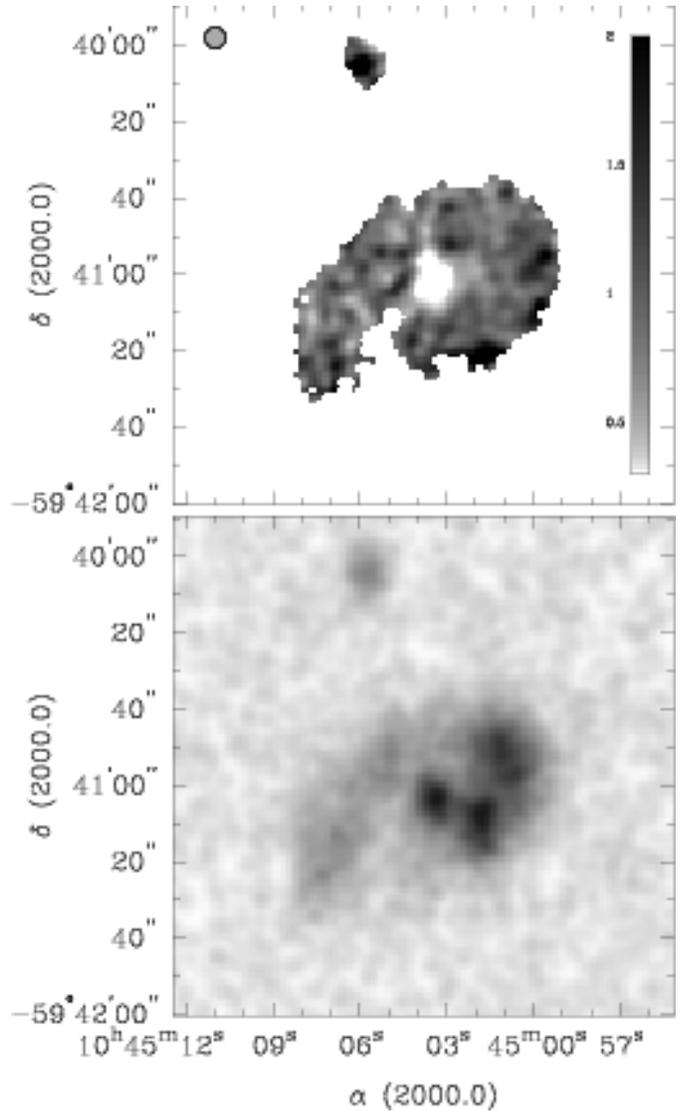}}}

\caption{The upper panel shows the ratio map between the 1994 and 1998
HRI data sets. Dark regions are brighter in the 1998 data. The lower panel
shows for comparison the image of the X-ray nebula. The effective resolution
of both images is plotted as filled circle in the upper panel.}
\label{fig:ratio}
\end{figure}

The newer, 1998, low state image (see Fig.\ \ref{fig:hrioo}, lower
panel) is better suited for such a comparison since the
signal-to-noise ratio (S/N) is comparable to the high state image.
Ratio images like Fig.\ \ref{fig:ratio} seem to imply that in
addition to some fluctuations of surface brightness, the region of
knot 2 brightened from 1992 to 1998. This effect may have created
the impression of expansion of the nebula as proposed by Corcoran
(priv. comm., and
http://lheawww.gsfc.nasa.gov/users/corcoran/eta\_car/
eta\_car.html\#HRI). The change in knot 1 seems not to be
confirmed by the 1998 data.  The faint emission at the position of
the central source seems to be present in both the 1992 and 1998
data sets.

This global expansion hypothesis by Corcoran reveals one major problem, it
predicts extremely high ($\sim 10\,000$ \kms) expansion velocities,
which are not found up to now. We tested several signal to noise
cuts and noted that in regions of low but still acceptable S/N
some brightening occurs. Since this effect is also visible at the
rim of the point source HD303308 (the star north of $\eta$ Car),
this may imply this rim brightening in the ratio images is an
artifact of low count rate and subsequent image processing.

Still there is an effect visible in Fig.\ \ref{fig:ratio}: as stated above,
the region of knot 2 seems to have brightened somewhat (factor of
about 1.6) in the 1998 data set compared to the 1994 data set
(and to lower confidence level compared to the 1992 data set).
Contrary to all the other differences between the images, which
are below the formal `beamsize' of the HRI and therefore not
significant, this effect is spatially extended and therefore, in
all likelihood, real.

\section{Comparison of the X-ray morphology, the kinematics and the HST
narrow band image}

As described in Sect.\ \ref{sect:xmorph}, the morphology of the
X-ray nebula and the optical appearance of the nebula around $\eta$ Carinae
are remarkably different. Except for $\eta$ Car itself and the
South Ridge, there seems to be no direct correlation between the
emission line image and the X-ray distribution (Fig.\
\ref{fig:hsthri}, top panel).

Our echelle spectra show a great deal of fine structure which
mainly results from numerous individual knots distributed all
over the observed region. In Fig.\ \ref{fig:echellograms} we give
examples for three different slit positions (Slit\,10S is located
10\arcsec\ south of our offset star, star \#60 in  Th\'e et al.\ 1980,
at a position angle of $132^\circ$, Slits\,38S and
42S are at 38\arcsec\ and 42\arcsec\ south,
respectively, at the same angle; for slit positions see Fig.\
\ref{fig:slits}). We have marked a number of
kinematically coherent structures. This was carried out for all
spectra; we then identified these structures with knots and
clumps on the HST image. Due to the much higher spatial
resolution of the HST in comparison with our spectra,
occasionally we found that structures that appear coherent in the
spectra consist of several smaller morphological entities.

In the lower panel of Fig.\ \ref{fig:hsthri} we give the maximum
radial velocities for individual knots and clumps. If in a small
area there are several knots with similar maximum radial
velocities, we give only the highest value of the entire
ensemble. To keep the crowding in the figure as small as
possible, negative velocities are underlined rather than marked
by a minus sign. In order to give the reader an easy overview, we
have binned the radial velocity ($v_{\rm rad}$) markings such
that we used four different font sizes, the smallest one for
$|v_{\rm rad}| \le 300\,$\kms\ and ever increasing ones for
$300\,$\kms $< |v_{\rm rad}| \le 600\,$\kms, $600\,$\kms $<
|v_{\rm rad}| \le 1000\,$\kms, and $1000\,$\kms $< |v_{\rm rad}|$.

A combination of the three pieces of information, morphology of
the HST image, radial velocities of the clumps and knots, and
morphology of the X-ray image clearly shows a good correlation
between the X-ray intensity and the absolute values of the radial
velocity (lower panel of Fig.\ \ref{fig:hsthri}). The most
intense regions in the X-ray image are those where one finds the
knots with the highest velocities. This, however, is usually not
a region where one finds strong emission in the optical (HST
image). We refer the reader to the above discussed positions, for
instance, where maxima of the absolute value of the radial
velocity and of the X-ray intensity coincide.

There is also a correlation between the lack of X-radiation and
the lack of material in the nebula around $\eta$ Car. A comparison
of Figs.\ \ref{fig:hrixray} and \ref{fig:hsthri} (lower panel)
shows that X-ray emission is missing south of $\eta$ Car (in the
area of $\alpha \approx 10^{\rm h} 45^{\rm m} 05^{\rm s}, \delta
\approx - 59\degr 42\arcmin 20\arcsec$), a region where there are
also less knots and clumps.

However, when comparing X-ray intensities and radial velocity
maxima, one has to keep in mind that

\begin{itemize}

\item radial velocities are good and bad tracers of the kinematics of the
material at the same time: While they give a lower limit of the
true three dimensional velocity of the material, they suffer from
the projection effect onto the line of sight. For motions almost
tangential to the celestial sphere, a thus determined lower limit
of the full velocity is no longer very meaningful. Unfortunately
outside of the Homunculus, in the outer ejecta only very few proper
motion measurements are available (Walborn et al.\ 1978, Walborn
\& Blanco 1988). For these measurements we find that the radial
velocities are in almost all cases comparable to or larger than
the tangential velocities, i.e., good tracers of the full
velocity field.

\item the highest velocities are not necessarily where the
shocks run into the densest clumps, which would be the dominant
producer of X-rays.

\item due to the lack of observable material---be it due to a true
local void, be it due to a too low brightness---the sampling of
the distribution of the radial velocities necessarily must be
less complete than that of the X-ray surface brightness.

\end{itemize}

The global distribution of the radial velocity maxima follows the
same orientation as the Homunculus, namely positive radial
velocities in the north-west and negative ones in the south-east.
We will discuss this in detail in a later paper (Weis 2000, Weis \&
Duschl, in prep.)

\section{Discussion}

Having established the spatial correlation between the peaks in
X-ray surface brightness of the nebula around $\eta$ Car with
regions of especially high radial velocity of the warm ionized
medium, we propose the following model for the X-ray surface
brightness distribution: material streaming away from $\eta$
Car---presumably having originated in the giant eruption of
1843---interacts with the ambient medium. This occurs with the
large relative velocities as indicated by the  radial velocity
measurements. The X-ray intensity is governed by the resulting
shock velocities and the densities. Due to the bandwidth of ROSAT
and in particular due to the high foreground H\,{\sc i} column
density ($\log N_{\rm H} \approx 21.3$, e.g., Savage et al.\ 1977)
only gas heated by shocks with velocities $\ge 300\kms$ is
observable. This implies that only high velocity shocked gas is
traced.

Knots 1 and 2 are the regions of highest X-ray surface brightness
in the X-ray nebula and coincide with regions with radial
velocities of about 1500 and 1900\kms , respectively. These high
velocities correspond to lower limits of the post-shock plasma
temperatures ($T_{\rm ps}$) of $3\,10^7$ and $5\,10^7$ K, with
$T_{\rm ps} = V^2 ( 3 \mu / 16 k)$ (see McKee 1987; $V$: radial
velocity taken as
representative for the shock velocity; $\mu$: mean mass per
particle). We assumed for these estimates a normal He/H ration of
0.1 leading to $\mu=0.61$ for a fully ionized gas.
Using a higher He/H ratio, as one would expect for the
ejecta of $\eta$ Car (increased He and N, decreased C and O as a
consequence of the CNO cycle), the resulting post-shock
temperatures would be even higher (e.g. by a factor of 1.5 for
He/H$=0.33$). These estimates unfortunately do not define
directly the overall intensity of the X-ray emission.  Such high
temperature gas has to be present, but it needs not to be
associated with high surface brightness.  To get a better handle
on the plasma temperature dominating the X-ray flux, we can
use the velocity structure of the X-ray knots:  The echelle spectra show
that clumps in this region with high surface brightness  move predominantly
with velocities of around 700\kms.  If the H$_{\alpha}$ and [N\,{\sc ii}]
surface brightness is a reliable tracer of the average gas density,
we can predict that the X-ray
plasma has a temperature of about $8\,10^6$ K, only somewhat
higher than the temperature of the soft component observed with
ASCA (Corcoran et al.\ 1998).

Over most of the area of the nebula the typical velocities are
smaller and about 400\kms\ to 600\kms, implying a lower
limit for the plasma temperature of $2-5\,10^6$\,K .

ASCA spectra of the $\eta$ Car X-ray nebula (Corcoran et al.\
1998) show that the emitting plasma can be well fitted with a two
component model, having plasma temperatures of $3.2\,10^6$ and
$6.3\,10^7$ K. The authors attributed the low temperature
component to the extended X-ray nebula and the hard component to
the central source ($\eta$ Car) itself. This idea is supported by
the limited information of the structure of the X-ray nebula
present in the PSPC data (Corcoran et al.\ 1995) for the eastern
part of the nebula. Interestingly, the temperature from the
integrated ASCA spectra fits well with the post-shock temperatures
derived above.  But there is an additional piece of evidence
concerning the properties of the X-ray nebula: during new ASCA
observation the central source was in low state. Corcoran et al.\
(2000) found that while the plasma temperatures of both the hard
and soft spectral component stays the same within the errors, the
absorbing H\,{\sc i} column densities of the hard spectral
component decreases together with the flux of the hard component.
Using these new ASCA data together with measurements at high
state, about 3\% to 6\% of the hard flux is still present, when
the central source is in low state (Corcoran et al.\ 2000). The
location of this hard emission is uncertain due to the large ASCA
PSF size, but recent CHANDRA data may support that the emission is
extended (see also Corcoran et al.\ 2000), and as already
suggested in Section \ref{sect:xmorph} . The lower limits of the
post-shock temperature implied from the radial velocities fit
perfectly to the measured temperature of the soft X-ray component.
Even more the presence of the very high velocity gas implies that
the extended nebula  at least contributes  partly to the flux in
the hard component.

The most distant regions of the nebula around $\eta$ Carinae show
expansion velocities of only 200\kms\ which would imply a
post-shock temperature of $5\,10^5$ K, too low to be detected with
the ROSAT HRI. Therefore, the sharp boundary of the X-ray nebula
in the north and east can be understood as an effect of the motion
of the nebula, too. In the south-west region lower surface
brightness wings from knot 1 and 2 extend into regions were no
high velocity optical emission is detected. This gas might be
missing because its optical emission is very faint and below the
detection limit of both the HST images and our echelle data. If
that is not the case some streaming of hot gas is needed to
explain the observations, instead of in situ creation of the hot
gas due to fast shocks passing through relatively dense gas.

A derivation of electron densities of the ambient medium from the
F656N filter HST image is not possible, as highly shifted
[N\,{\sc ii}] emission contaminates the H$_\alpha$ band. While it
is tempting to derive electron densities and the thermal energy
content (e.g. Chu \& MacLow 1990) of the knots 1 and 2, we
decided against this exercise, since the correction factors from
the ROSAT HRI fluxes in the 0.1 to 2.4\,keV band to the
approximate temperatures of the X-ray gas in the knots of at
least $3\,10^7$\,K would be too large. If we assume that the
typical temperature of the knots is defined by the velocity of
the majority of the structures in the knot 1 and 2 (about 700\kms),
rms electron densities could be derived. But still, due to the
complexity of the
shape of the X-ray emitting knots and the fact that they consist
of multitudes of different clumps moving at different velocities,
the values would be almost meaningless for the physical state of
the emitting gas.  Additionally, the temperature of the hot gas
is derived for thermal equilibrium (using Raymond-Smith plasma
models), while the topology of the emitting region and the small
time scales appropriate for $\eta$ Carinae (of the order of 100 years!)
implies strongly non-equilibrium conditions. High spatial
resolution and high signal to noise spectra as obtainable with
the CHANDRA satellite may be able to probe this and derive
meaningful physical parameters for the X-ray emitting plasma in
the $\eta$ Carinae nebula.  Still, it is interesting to note, that the
X-ray surface brightness in the ROSAT energy band of knot 1 and 2
are comparable or even larger than that of the central source
itself.

If the contribution of the nebula to the hard X-ray emission of
the nebula around $\eta$ Car is relatively small, as implied by
the ROSAT PSPC  results (Corcoran et al.\ 1995) and the new
CHANDRA data (Seward et al. 2001), we are left with a problem: why
are pronounced bright spots of the relatively soft X-ray emission,
namely knot 1 and 2, at the exact position of the fastest moving
structures seen in our echelle spectra?  The X-ray emission does
not correlate well with the surface brightness of the ionized gas
or even the surface brightness or number density of the knots
moving faster than $300\kms$, which are able to produce observable
X-ray emission of the $\eta$ Carinae nebula. If the X-ray emission
at these knots is not dominated by hard X-ray emission produced by
the very fast shocks implied by our velocity measurements, as the
early release CHANDRA data imply, we need another process to
explain this tight correlation.

One way out could be the following:  If for the gas the time
after the passage of the fast shock is too short to reach thermal
equilibrium, the kinetic temperature of the gas is high, while
the adaption of ionization state runs ahead. Such non-equilibrium
effects were discussed by e.g. Schmutzler \& Tscharnuter (1993).
Without knowledge of the exact ionization history we can only
check the feasibility of this idea using time-scales. We use here
Nitrogen as a trace element, since the normally dominant coolant
Oxygen is strongly suppressed in the ejecta nebula of  $\eta$ Car
due to the CNO cycle (Davidson et al.\ 1982, Dufour et al.\ 1997).
At $5\,10^7$\,K the dominant ion of Nitrogen is N\,{\sc vi}
(Sutherland \& Dopita 1993). Using the recombination coefficient
calculated with the tabulations of Shull \& van Steenberg (1982)
and a typical density in the outer ejecta (more precise the S
ridge) of at least $n_{\rm e} \sim 12\,000$\,cm$^{-3}$ (Dufour et
al.\ 1997) the recombination time is $\sim 50$ yrs.   Using the
cooling coefficients tabulated by Sutherland \& Dopita (1993) and
the same density the cooling time of this plasma is 1400 yrs. We
would therefore observe a plasma which appears colder when
measured from its ionization state, than it is according to its
internal state. The net effect is that an analysis with an
equilibrium plasma code would derive a too low temperature, since
the line emission and therefore the ionization conditions dominate
the spectrum. This may be the case for the knots 1 and 2 and could
present the solution for the otherwise strange correlation of the
highest velocities with peaks of relatively low temperature X-ray
plasma traced by the ROSAT HRI images.  Most other regions of the
nebula would be near equilibrium conditions as indicated by the
close match of the X-ray temperature and the post-shock
temperatures derived from our high resolution echelle spectra.

Again, to test this
idea, high spatial and spectral resolution X-ray spectra with
good signal to noise are needed, which may be provided by
CHANDRA observations.

\section{Summary}

A comparison of the morphological appearance of the $\eta$ Car
nebula in the optical (HST F656N filter) and the X-ray (ROSAT
HRI) wavelengths ranges shows remarkable differences. Adding as a
third piece of information the kinematics of the material, one
finds that there is a strong correlation between the largest
absolute values of the radial velocities of knots and clumps and
the areas of the highest X-ray surface brightnesses. This offers
as an explanation that the X-radiation as traced by the HRI
images comes predominantly from regions where material streaming
away from $\eta$ Car interacts with the ambient medium and gives
rise to shock fronts. The observed velocities allow us to derive
a post-shock temperature. The resulting value is in good
agreement with temperatures derived from ASCA spectra, but is
contradicted to the limited spatial information provided by the
ROSAT PSPC.  While some contribution of the nebula  the hard
emission is still possible, the problem is to reconcile the
strong correlation of the region of highest velocity (not surface
brightness) ionized gas with the strongest X-ray peaks.  With the
current data non-equilibrium ionization appears the only way out,
but better spectra are highly needed to test the idea.  In
summary, one can say that while the morphology of the optical and
the X-ray images are rather different, the underlying physical
process, i.e., the interaction between the material streaming
away from $\eta$ Car and the ambient medium yields the clue to
the relation between the images in the two wavelengths regimes.

\begin{acknowledgements}
KW is greatful to Prof.\ You-Hua Chu for motivation and discussions
on the subject of this paper. DJB thanks the ITA director
Prof.\ Dr.\ W.M.\  Tscharnuter and the ITA staff for
their great hospitality during the several visits while working
on this paper. The authors thank the referee, Dr. M.\ Corcoran,
for many comments and suggestions which helped to improve the
paper considerably. The data reduction and analysis was in part
carried out on a workstation provided by the {\it Alfried Krupp
von Bohlen und Halbach Stiftung\/}. Guest User, Canadian Astronomy
Data Centre, operated by the Herzberg Institute of Astrophysics,
National Research Council of Canada. Data retrieved from the
ROSAT archive at MPE Garching.

\end{acknowledgements}

\end{document}